\documentclass[9pt,conference]{IEEEtran}
\IEEEoverridecommandlockouts
\usepackage{cite}
\usepackage{amsmath,amssymb,amsfonts,mathtools}
\usepackage{graphicx}
\usepackage{textcomp}
\usepackage{booktabs}
\usepackage{multirow}
\usepackage{listings}
\usepackage{adjustbox}
\usepackage{makecell}
\usepackage{cite}
\usepackage{bigdelim}
\usepackage{algorithmic}
\usepackage{algorithm}
\usepackage{array}
\usepackage{textcomp}
\usepackage{stfloats}
\usepackage{url}
\usepackage{verbatim}
\usepackage{diagbox}
\usepackage{balance}
\usepackage{tabularx}
\usepackage{color,soul}
\usepackage{hyperref}
\usepackage{tabu}
\usepackage{flushend}
\usepackage[normalem]{ulem}

\usepackage[T1]{fontenc}

\usepackage[labelformat=simple]{subcaption}

\usepackage[font=small,labelfont=bf]{caption}

\usepackage{titlesec}
\titlespacing*{\section}{0pt}{0.2\baselineskip}{0.1\baselineskip}
\titlespacing*{\subsection}{0pt}{0.2\baselineskip}{0.1\baselineskip}
\titlespacing*{\subsubsection}{0pt}{0.1\baselineskip}{0.0\baselineskip}

\renewenvironment{thebibliography}[1]{
  \begin{oldthebibliography}{#1}
    \setlength{\itemsep}{0em}
    \setlength{\parskip}{0em}
}
{
  \end{oldthebibliography}
}
\usepackage{setspace}

\allowdisplaybreaks

\usepackage[T1]{fontenc}
\def\BibTeX{{\rm B\kern-.05em{\sc i\kern-.025em b}\kern-.08em
    T\kern-.1667em\lower.7ex\hbox{E}\kern-.125emX}}

\usepackage{pifont}
\newcommand{\cmark}{\ding{51}}%
\newcommand{\xmark}{\ding{55}}%
    
\begin{document}

\title{M-BEST-RQ: A Multi-Channel Speech Foundation Model for Smart Glasses}

\author{\begin{tabular}{c}Yufeng Yang$^{1*, 2}$\thanks{*Work done during internship at Meta}, Desh Raj$^1$, Ju Lin$^1$, Niko Moritz$^1$, Junteng Jia$^1$, Gil Keren$^1$, \\ 
Egor Lakomkin$^1$, Yiteng Huang$^1$, Jacob Donley$^1$, Jay Mahadeokar$^1$, Ozlem Kalinli$^1$\end{tabular}\\
$^1$Meta, USA\quad $^2$The Ohio State University, USA\\
\texttt{yang.5662@osu.edu, desh@meta.com}
}
\maketitle

\begin{abstract}
The growing popularity of multi-channel wearable devices, such as smart glasses, has led to a surge of applications such as targeted speech recognition and enhanced hearing. However, current approaches to solve these tasks use independently trained models, which may not benefit from large amounts of unlabeled data. In this paper, we propose M-BEST-RQ, the first multi-channel speech foundation model for smart glasses, which is designed to leverage large-scale self-supervised learning (SSL) in an array-geometry agnostic approach. While prior work on multi-channel speech SSL only evaluated on simulated settings, we curate a suite of real downstream tasks to evaluate our model, namely (i) conversational automatic speech recognition (ASR), (ii) spherical active source localization, and (iii) glasses wearer voice activity detection, which are sourced from the MMCSG and EasyCom datasets. We show that a general-purpose M-BEST-RQ encoder is able to match or surpass supervised models across all tasks. For the conversational ASR task in particular, using only 8 hours of labeled speech, our model outperforms a supervised ASR baseline that is trained on 2000 hours of labeled data, which demonstrates the effectiveness of our approach.

\end{abstract}


\section{Introduction}
With the growing popularity and adoption of multi-channel smart wearable devices such as smart glasses, there are several new use cases related to the spatial understanding of audio and speech. These include automatic speech recognition (ASR) for smart assistants, enhanced hearing, etc~\cite{engel2023project}. Smart wearable devices usually consist of multi-channel audio input and involve the wearer’s interaction with the device and surrounding objects or participants. However, because of the limited annotated data from such devices, the use of multi-channel inputs is often limited to traditional signal processing. Furthermore, different use cases are usually addressed using separate models that do not make use of the knowledge from other tasks. Self-supervised learning (SSL) has been shown to be effective on low-resource tasks with representations learned from unlabeled data~\cite{mohamed2022self, schneider2019wav2vec, baevski2020wav2vec, hsu2021hubert, chen2022wavlm, Yang2021SUPERBSP}, and ``foundation models'' trained using such methods have recently outperformed supervised models~\cite{Bommasani2021OnTO}. Framing the above problem in the low-resource context, our objective in this work is to build the first foundation model specifically for tasks based around wearable devices such as smart glasses.

Most existing work on speech SSL has focused on single-channel inputs~\cite{schneider2019wav2vec, baevski2020wav2vec, hsu2021hubert, chen2022wavlm}. For multi-channel SSL, while researchers have proposed methods such as Spatial HuBERT~\cite{dimitriadis2023spatial}, multi-channel AV-wav2vec2~\cite{zhu2024multichannel}, and UniX-Encoder~\cite{huang2024unix}, these models have only been evaluated in limited settings of simulated data or fixed array-geometry. Different from these, we want to build a foundation model that be fine-tuned on several downstream tasks and can work across wearable devices with different numbers of microphones and array geometries. Our key insight to achieve device-agnosticity is to use multiple super-directivity beamformers to convert ``channels'' to a fixed number of ``directions'' which can be processed by the neural encoder~\cite{lin2023directional}. For the SSL backend, we propose a \underline{\smash{multi-channel extension of BEST-RQ}}~\cite{chiu2022bestrq}, since it is conceptually simple and has been shown to outperform other methods such as wav2vec 2.0~\cite{whetten2024open}. We refer to the resulting model as \textbf{M-BEST-RQ}.

Following the conventional paradigms, we first pre-train the M-BEST-RQ encoder using masked estimation methods on a combination of large-scale synthetic data simulated from public datasets (such as LibriSpeech~\cite{panayotov2015librispeech} and Libri-Light~\cite{kahn2020librilight}) and real multi-channel data from the Project Aria glasses~\cite{engel2023project}. We then use the encoder for supervised fine-tuning and evaluation on several downstream tasks: conversational ASR (C-ASR), spherical active source localization (S-ASL), and glasses wearer voice activity detection (W-VAD). On the C-ASR task (formulated using the recently published MMCSG dataset~\cite{zmolikova2024chime}), M-BEST-RQ achieves 20.1\%/28.1\% word error rate (WER) for self/other speaker using only 8 hours of labeled speech, outperforming an ASR baseline trained on 2k hours labeled data. We curate the S-ASL and W-VAD tasks using the EasyCom dataset~\cite{donley2021easycom}, which is recorded on a different device than Aria. On these tasks, our audio-only M-BEST-RQ model matches or outperforms baselines trained with audio-visual modalities, indicating that M-BEST-RQ is a generic foundation model that can work for several downstream tasks on different devices.


\section{M-BEST-RQ}
\label{sec:method}

\begin{figure*}[!htbp]
    \centering
    \includegraphics[width=0.8\linewidth,trim={1cm 0 0 0},clip]{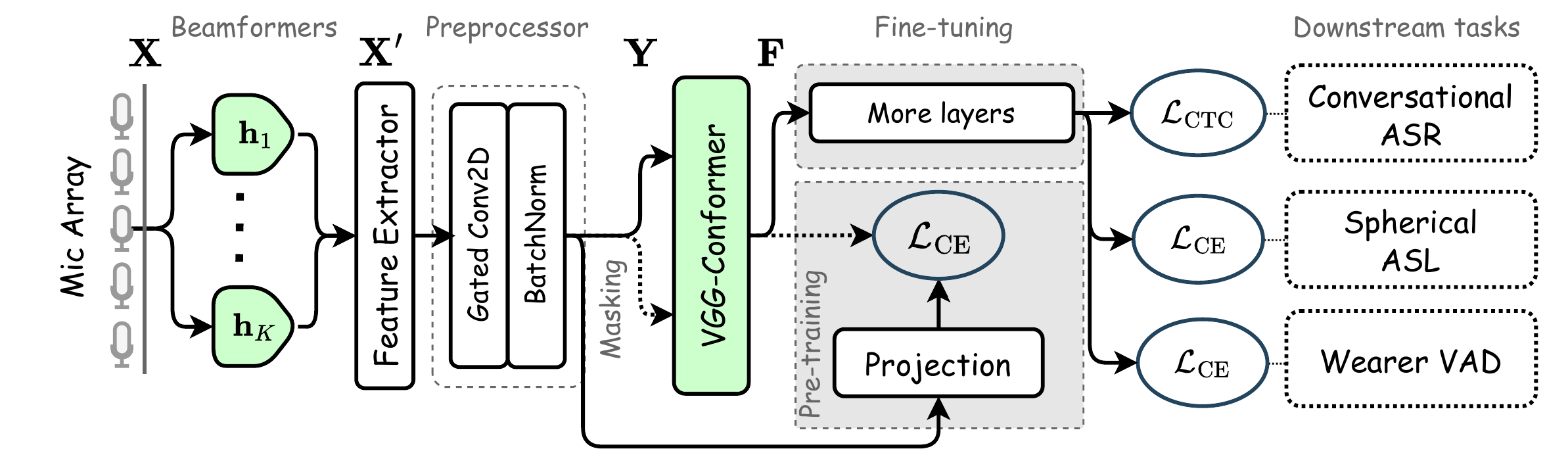}
    \caption{System architecture of M-BEST-RQ and downstream tasks.}
    \label{fig:m_best_rq_arch}
\end{figure*}

Given raw speech $\mathbf{X}\in \mathbb{R}^{T\times M}$ containing $T$ samples collected on $M$ microphones, the problem is to learn a function $f$ (the foundation model) which converts $\mathbf{X}$ into task-and-array-agnostic high-dimensional representations $\mathbf{F}^{T'\times D}$, where $T'$ is usually down-sampled from $T$. At a high-level, there are two questions that must be answered in order to achieve this: (i) How do we make the foundation model $f$ invariant to the number of microphone channels $M$ and their geometry? (ii) How do we learn $f$ such that it is ``generic,'' i.e., it can perform well on any downstream task $\mathcal{T}$?

\subsection{Array invariance with fixed beamformers}

For general-purpose multi-channel devices, array geometry invariance may be achieved using neural methods such as cross-channel attention~\cite{Chang2021EndtoEndMT} at the cost of increased computation. However, this solution disregards the fact that all the array geometries, despite being different, are situated on wearable devices and are used for wearer-related tasks. With this assumption in place, we can instead make use of device-specific beamformers to convert an arbitrary number of input ``channels'' $M$ to a fixed number of input ``directions'' $K$, as shown in Fig.~\ref{fig:m_best_rq_arch}. Formally, we have
\begin{equation}
    \mathbf{F} = f(\mathbf{X}) = (f_{\mathrm{enc}}\circ f_{\mathrm{bf}})(\mathbf{X}) = f_{\mathrm{enc}}\left(f_{\mathrm{bf}}(\mathbf{X})\right),
\end{equation}
where $f_{\mathrm{bf}}(\mathbf{X}) = \mathbf{X}' \in \mathbb{R}^{T\times K}$ is fixed, and $f_{\mathrm{enc}}$ is parameterized by a neural network. Here, $\mathbf{X}'$ is invariant to the array geometry, and represents a collection of signals from $K$ directions. 

For our function $f_{\mathrm{bf}}$, we use non-linearly constrained minimum-variance (NLCMV)
beamforming~\cite{lin2023directional, lin2024agadir}. Given the mouth-directed and far-field acoustic transfer functions (ATFs), the beamforming weights $\mathbf{h}_k(j\omega)$ of each steering direction $k \in \{1,\ldots,K\}$ are obtained by minimizing
\begin{equation}
    \mathbf{h}^{H}(j\omega) \left[ \mathbf{\Phi}_{dd}(j\omega) + \underbrace{\phi_{pp}(\omega)\sum_{n=1}^{N}\alpha_{p,n}\cdot \mathbf{g}_{n}(j\omega)\mathbf{g}_{n}^{H}(j\omega)}_{\text{soft control of null directions.}} \right] \mathbf{h}(j\omega),
\end{equation}
which is subject to the linear equality and nonlinear inequality constraints, and they are simplified to
\begin{equation}
\left\{
\begin{aligned}
     & \mathbf{h}^{H}(j\omega)\mathbf{g}(j\omega) = 1,\\
     & c(\omega)\triangleq \underbrace{\mathbf{h}^{H}(j\omega) \mathbf{\Psi}(j\omega) \mathbf{h}(j\omega) \leq 0}_{\text{constraint on white noise gain.}},
\end{aligned}
\right.
\end{equation}
where $\mathbf{\Phi}_{dd}(j\omega)$ is the covariance matrix of diffuse noise, and
\begin{equation}
    \mathbf{\Psi}(j\omega) \triangleq \mathbf{I} - \mathbf{g}_{n}(j\omega)\mathbf{g}_{n}^{H}(j\omega)\cdot M \Bigg / \left[ \sum_{m=1}^{M} |G_{m}(j\omega)|^{2} \right].
\end{equation}

The $G_{m}(j\omega)$ are measured channel responses from the target speech source to the $m$-th microphone (ATFs). $N$ is the number of point noise sources, $\phi_{pp}(\omega)$ is the power spectral density of point noise, $\alpha_{p,n}$ is the $n$-th point noise weight, and $\mathbf{I}$ is the identity matrix. The nature of the directive signal enables the model to learn array-geometry agnostic representations, and with ATFs available, the NLCMV beamforming can be applied to any new device, i.e., $\mathbf{X}' = \mathbf{h}_k(j\omega)\mathbf{X}$.

\subsection{Task invariance with BEST-RQ}

We extend the BEST-RQ~\cite{chiu2022bestrq} to work with multi-channel inputs by affixing a multi-channel projection module in front. Given $\mathbf{X}'$, we extract the corresponding log-Mel filterbank features, and then project the $K$ channels into a single channel using a gated convolution followed by batch-normalization, thus resulting in latent representations $\mathbf{Y}$ which contain information from all $K$ directions. These representations are provided as input to a VGG-Conformer encoder ($f_{\mathrm{enc}}$), which outputs semantic embeddings $\mathbf{F}$. 

Similar to \cite{chiu2022bestrq}, we train the encoder by masking random chunks of $\mathbf{Y}$ before feeding to $f_{\mathrm{enc}}$. An unmasked $\mathbf{Y}$ is projected using a randomly initialized quantizer into discrete labels, and the pre-training objective is to predict the labels corresponding to the masked regions using cross-entropy loss, $\mathcal{L}_{\mathrm{CE}}$. During the fine-tuning stage, we add additional layers to the output of the VGG-Conformer encoder and train it on labeled data for different downstream tasks with different loss functions, as shown in Fig.~\ref{fig:m_best_rq_arch}. Our conjecture is that the masking $\mathbf{Y}$ in the pre-training stage imparts semantic as well as directional understanding of multi-channel speech to the M-BEST-RQ encoder, which would enable it to work well on several tasks. 

\section{Experimental Setup}
\label{sec:exp}

\subsection{Datasets}
We simulated 7-channel LibriSpeech (LS) and Libri-Light (LL) datasets according to the array configuration of the Project Aria glasses, based on the original LibriSpeech~\cite{panayotov2015librispeech} and Libri-Light~\cite{kahn2020librilight} datasets. For this, we first segmented the long utterances into shorter segments (between 0.5 and 10 seconds) based on forced alignments\footnote{Since Libri-Light does not contain transcriptions for each utterance, we used an in-house ASR model to get pseudo-labels for simulation purposes.}. We then generated 100k room impulse responses (RIRs) for the Aria microphone array, and used these to simulate multi-channel, multi-speaker conversations between a wearer, a participant, and a distractor speaker. The simulation process is the same as the ``train-from-scratch'' baseline of the MMCSG challenge~\cite{zmolikova2024chime}. The duration of the simulated multi-channel LS and LL datasets are about 2k and 140k hours, respectively, with each utterance being 12$\pm$5 seconds.

In addition, we also used $\sim$800 hours of real, in-house, multi-channel data collected using the Aria glasses to investigate the impact of real data in pre-training. We downsampled these recordings from 48 kHz into 16 kHz and randomly segmented the recording into 12$\pm$5 second segments, resulting in segments in the range [2, 30] seconds. We refer to this dataset as RD.

\begin{figure}[t]
    \centering
    \begin{subfigure}[h]{0.5\linewidth}
        \centering
        \includegraphics[width=\linewidth]{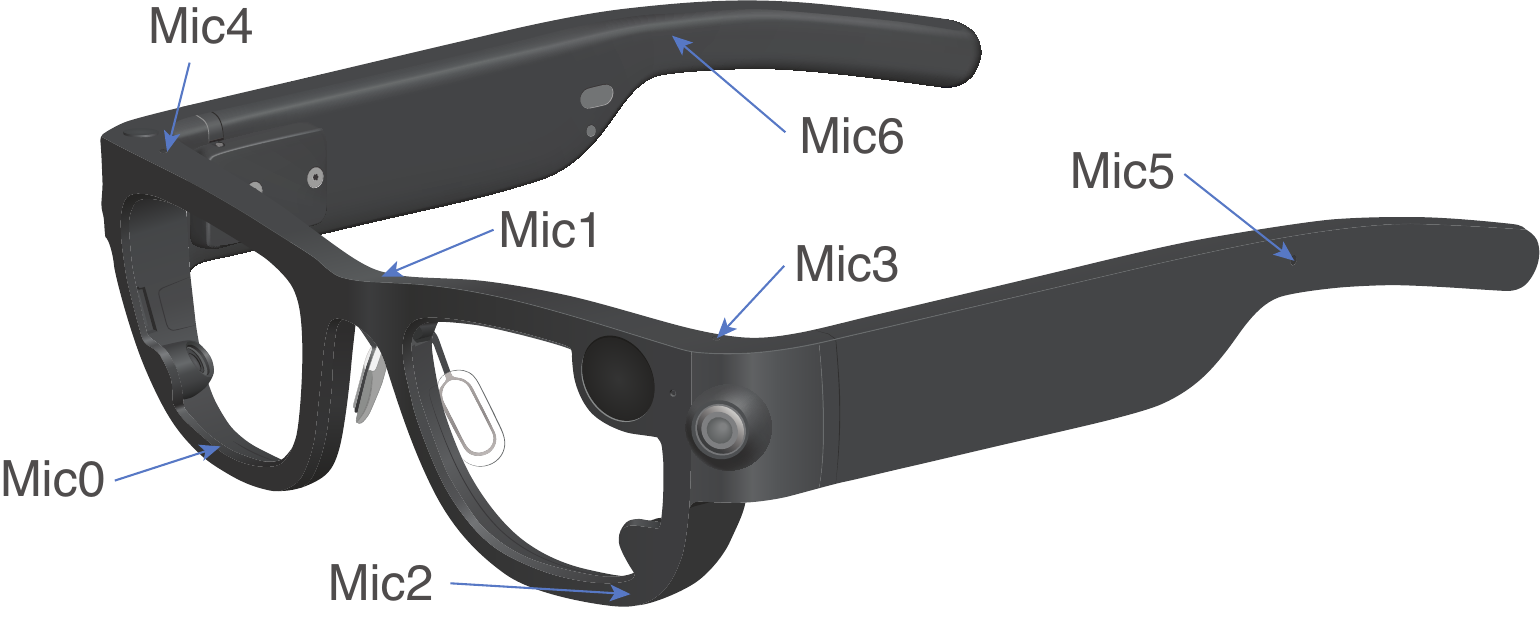}
        \caption{Project Aria glasses}
        \label{fig:aria}
    \end{subfigure}
    \hfill
    \begin{subfigure}[h]{0.45\linewidth}
        \centering
        \includegraphics[width=\linewidth]{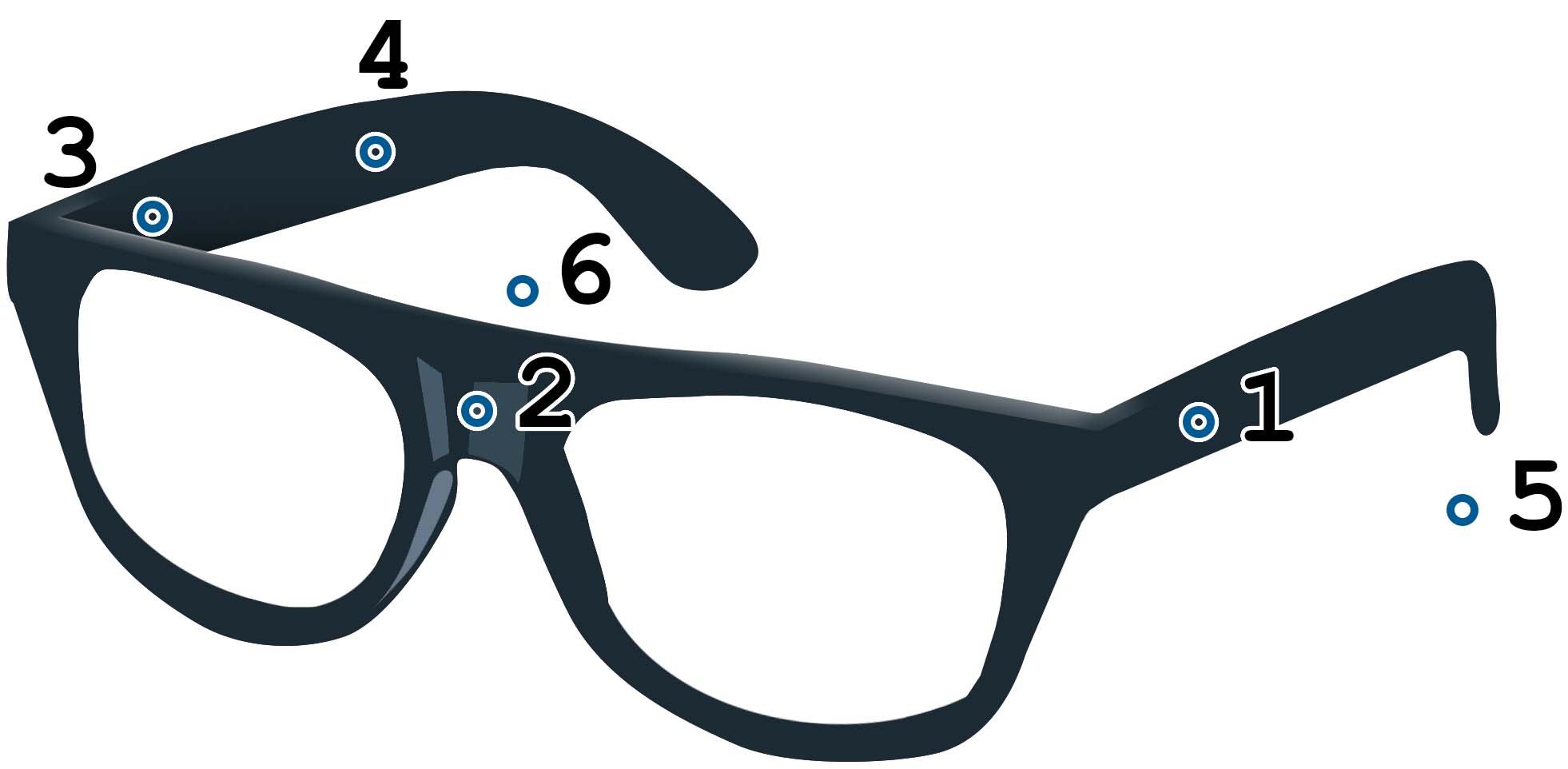}
        \caption{EasyCom AR glasses}
        \label{fig:meta_ar}
    \end{subfigure}
    \caption{Microphone positions of two devices: (a) Aria glasses, containing 7-channel input, and (b) EasyCom AR glasses, containing 6-channel input. For (b), we only used the first 4 microphones which are on the device.}
    \label{fig:glasses_mic}
\end{figure}

For fine-tuning and evaluation, we curated 3 downstream tasks based on the MMCSG~\cite{zmolikova2024chime} and EasyCom~\cite{donley2021easycom} datasets. MMCSG is released as part of the CHiME-8 challenge\footnote{\url{https://www.chimechallenge.org/current/task3}} focusing on transcribing natural conversations between two speakers, recorded on Aria glasses. The duration of training, development, and evaluation data is 8.5, 8.4, and 9.4 hours, respectively. EasyCom is a dataset of multi-talker conversations in noisy environments with egocentric video recorded on an augmented-reality (AR) glasses with a different number and array of microphones from the Aria glasses. The duration of the dataset is about 5.3 hours. Details of the microphone positions of two glasses are shown in Fig.~\ref{fig:glasses_mic}. For the EasyCom AR glasses, we only used the first 4 microphones which are on the device. Since EasyCom has an audio/video frame rate of 20 Hz, while the frame rate of our M-BEST-RQ encoder is 25 Hz, we resampled the frames in EasyCom by 0.8 through repetition and subsampling.

\subsection{Pre-training}
\label{sec:pre-train}

We trained four M-BEST-RQ models on different combinations of datasets: LS, RD, LS+RD, and LL. All models share the same architecture, containing fixed beamformers, a log-mel filterbank extractor, gated 2-D convolutions, and a VGG-conformer encoder. The VGG-Conformer encoder consists of 2 VGG~\cite{simonyan2015vgg} subsampling layers and 24 Conformer encoder layers~\cite{conformer} with a hidden dimension of 512. The number of trainable parameters of M-BEST-RQ is $\sim$96M. We used 32 NVIDIA A100 GPUs to train M-BEST-RQ on LS, RD, and LS+RD, and 128 A100 GPUs to train M-BEST-RQ on LL. For all pre-training, the Adam optimizer was used with betas of (0.9, 0.98), epsilon of $1e^{-8}$, and weight decay of $1e^{-4}$. A tri-stage learning rate schedule was used with a peak learning rate of 0.0003, warmed up for 20k steps (or 30k for LL pre-training). We used 2048 code-books, each of size 24, for the random quantizer. The mask probability and lengths were set to 0.02 and 30 frames, respectively, based on our preliminary investigation. For fine-tuning, we selected models from different checkpoints for each of LS (600k steps), RD (400k steps), LS+RD (600k steps), and LL (850k steps), based on the convergence of the training accuracy of codebooks.

\subsection{Downstream Tasks}
As mentioned earlier, most existing work on multi-channel SSL has conducted evaluations on simulated tasks. In order to evaluate M-BEST-RQ on real settings, we curated three downstream tasks focused on the smart glasses use case. We describe these tasks and their implementation details below.

\subsubsection{Conversational ASR}
The C-ASR task is based on CHiME-8 Task-3, using the MMCSG dataset. It consists of conversations recorded between the wearer and a participant (the speaker located directely in front of the wearer) in the presence of background noise and distracting speakers. The task objective is to transcribe and attribute the speech from both the wearer (self) and the participant (other). Performance is measured using the speaker-attributed WER metric. We used the same data preparation process as the official baseline system for MMCSG. This involves cropping each recording to $\sim$20 second segments and using serialized output training (SOT) transcripts \cite{kanda2020serialized}. We inserted \guillemotright$0$ and \guillemotright$1$ before each token in the reference to indicate whether the token is attributed to self or other, respectively.

For M-BEST-RQ fine-tuning, we added a 4096-dimensional linear head (one for each sentence-piece including \guillemotright$0$ and \guillemotright$1$) on the output of the VGG-Conformer encoder and trained with CTC loss~\cite{graves2006connectionist}. We used 8 A100 GPUs during fine-tuning with batch size 256. We warmed up the learning rate to $3e^{-5}$ for 6000 warm-up steps and then decayed it exponentially. During fine-tuning, we used real volume perturbation to scale the volume of conversations within the range from 0.01 to 0.99. We also added a SpecAugment layer \cite{park2019specaugment} after the feature extraction.

Since the official challenge baseline is a streaming RNN-T model~\cite{noroozi2024stateful}, we prepared a comparable ASR baseline which shares the same model architecture as our fine-tuned M-BEST-RQ. We first trained this ASR model on LS, and then fine-tune on MMCSG train set after convergence.

\subsubsection{Spherical Active Source Localization}\label{sec:asl}
The presence of multiple microphones on smart glasses enables localization for the active source around the wearer. S-ASL~\cite{jiang2022egocentric} predicts a (90, 180) feature map where 90 and 180 denote the elevation and azimuthal, respectively, with a 2° resolution. The position in the feature map indicates the angles of directions and is computed by transferring the annotation of 3D points and quaternions to the (90, 180) map where 0/1 indicate the absence/presence of a speech source. Following \cite{zhao2023audio}, we treat this task as classification and follow the same step in terms of training and evaluation. During fine-tuning, we added two linear layers at the output of the VGG-Conformer encoder, projecting the hidden dimension to 4050. After reshaping to (45, 90), two tensors are upsampled and concatenated to (2, 90, 180). We apply the same non-maximum suppression of radius 5 and threshold 0 and match the augmented ground truth with the Hungarian algorithm. The evaluation metrics are mean angular errors (MAE) for the distance from prediction to ground truth (indicating false positives), and from ground truth to prediction (indicating missing targets). Training was done on 16 A100 GPUs with a batch size of 64, and a tri-stage learning rate with 2500 warm-up and 17500 decay steps, with a peak learning rate of $3e^{-5}$. Same as \cite{jiang2022egocentric, zhao2023audio}, we used sessions 4-12 for training and 1-3 for testing on EasyCom. 

\subsubsection{Glasses Wearer VAD}\label{sec:vad}
We define this task as frame-level binary classification where for each frame, the model output is 1 if the glasses wearer is speaking and 0 otherwise. We used the same EasyCom dataset as in S-ASL to create this task. Following \cite{jiang2022egocentric}, for the fine-tuning of M-BEST-RQ, we added two linear layers at the end of the VGG-Conformer encoder to perform binary classification. The training was done on 8 A100 GPUs with a batch size 128, with a peak learning rate of $3e^{-5}$ warmed up for 3000 steps. As before, we used sessions 4-12 for training and 1-3 for testing. We computed the mean average precision (mAP) as the metric to evaluate each model.

\begin{table*}[t]
    \centering
    \caption{C-ASR results on the MMCSG evaluation set, in terms of self and other WER (\%). We also report WER breakdown into insertion (\texttt{ins}), deletion (\texttt{del}), substitution (\texttt{sub}), and speaker attribution (\texttt{attr}) errors. $\dagger$ denotes the official challenge baseline. \underline{Underline} denotes labeled data in pre-training. All fine-tune data are labeled.}
    \label{tab:mmcsg_wer}
    \centering
    \resizebox{0.8\linewidth}{!}{
    \begin{tabular}[width=\linewidth]{@{}llrrrrrrr@{\hskip 0.25in}rrrrr@{}}
        \toprule
         \multirow{2}{*}{\textbf{Model}} & \multicolumn{2}{c}{\textbf{Pre-train}} & \multicolumn{1}{c}{\textbf{Fine-tune}} & \multicolumn{5}{c}{\textbf{Self}} & \multicolumn{5}{c}{\textbf{Other}} \\
         \cmidrule(l{0.5em}r{0.5em}){2-3} \cmidrule(l{0.5em}r{0.5em}){4-4} \cmidrule(l{0.4em}r{0.25in}){5-9} \cmidrule{10-14}  
            & Data & Size (h) & Size (h) & \textbf{WER} & \texttt{ins} &  \texttt{del} & \texttt{sub}  & \texttt{attr} & \textbf{WER} & \texttt{ins} &  \texttt{del} & \texttt{sub}  & \texttt{attr} \\
        \midrule
        RNN-T$^\dagger$ & LS++ & \underline{4.5k} & 8 & 22.0 & 2.7 & 4.3 & 13.5 & 1.6 & 32.8 & 4.2 & 8.0 & 17.9 & 2.6 \\
        \midrule
        \multirow{2}{*}{CTC} & None & 0 & 8 & 67.8 & 4.0 & 15.2 & 46.2 & 2.4 & 76.4 & 4.8 & 14.9 & 52.1 & 4.5 \\
         & LS & \underline{2k} & 8 & 22.9 & 2.5 & 4.3 & 14.8 & 1.4 & 30.8 & 3.7 & 7.3 & 17.9 & 1.9 \\
        \midrule
        \multirow{6}{*}{\textbf{M-BEST-RQ}} & (A) RD & 800 & 8 & 40.5 & 4.0 & 7.2 & 28.0 & 1.4 & 49.6 & 5.3 & 9.4 & 32.7 & 2.2 \\
         & (B) LS & 2k & 8 & 24.3 & 2.2 & 4.1 & 16.6 & 1.4 & 33.4 & 3.8 & 7.2 & 20.5 & 1.8 \\
         & (C) LS+RD & 2.8k & 8 & 21.5 & \textbf{2.0} & 3.7 & 14.3 & 1.4 & 29.5 & \textbf{3.5} & 6.7 & 17.9  & 1.4 \\
         & (D) LL  & 140k & 8 & \textbf{20.1} & \textbf{2.0} & \textbf{3.5} & \textbf{13.4} & \textbf{1.2} & \textbf{28.1} & 3.7 & \textbf{6.2} & \textbf{16.8} & \textbf{1.3} \\
         \cmidrule{2-14}
         & (E) LS+RD & 2.8k & 2k+8 & 16.7 & 1.8 & \textbf{3.2} & 10.5 & \textbf{1.1} & 24.5 & 3.3 & \textbf{6.2} & 13.9 & \textbf{1.1}\\
         & (F) LL & 140k & 2k+8 & \textbf{16.5} & \textbf{1.6} & 3.6 & \textbf{10.2} & \textbf{1.1}  & \textbf{23.8} & \textbf{3.0} & 6.7 & \textbf{12.9} & 1.3 \\
        \bottomrule
    \end{tabular}}
\vspace{-1em}
\end{table*}

\section{Results \& Discussion}
\label{sec:result}

\subsection{Conversational ASR}
We compare the results of different systems in Table~\ref{tab:mmcsg_wer}. The challenge baseline is an RNN-T ASR model \cite{noroozi2024stateful} which has $\sim$114 M parameters, and is pre-trained on 4.5k hours of 7-ch perturbed LS and TED-LIUM~\cite{hernandez2018ted} (denoted as LS++). We reproduce the official baseline results here~\cite{zmolikova2024chime}. All other systems are models trained with CTC loss with $\sim$98 M parameters. For the CTC models, we included a system trained directly on the MMCSG training set in addition to our ASR baseline. Systems (A), (B), (C), and (D) use M-BEST-RQ encoders pre-trained on RD, LS, LS+RD, and LL, respectively. We also included systems (E) and (F), which use LS for fine-tuning, in addition to MMCSG.

First, we see that our CTC-based ASR baseline is competitive with the challenge baseline, despite using less pre-training data. Among the M-BEST-RQ systems fine-tuned only with 8h of MMCSG data, systems (A) and (B), trained on RD and LS, respectively, were found to be worse than the ASR baseline. However, system (C), which was trained on LS+RD, achieved 21.5\%/29.5\% WER on self/other speaker, outperforming the ASR baseline by over 1\% absolute WER reduction. This indicates that pre-training using a combination of synthetic and real data may be important for the M-BEST-RQ model, when the size of synthetic data is small. Nevertheless, if large-scale simulated data is used, real data may not be required, as shown by the strong performance of system (D). This system, despite being pre-trained on LL only, achieved 20.1\%/28.1\% WER on self/other, outperforming the ASR baseline by 2\%. Systems (E) and (F) further improved the WER by over 6\% absolute, demonstrating the importance of labeled data.

\subsection{Spherical Active Source Localization}
\label{result:asl}

Following \cite{yun2024spherical}, we compared the MAE from prediction to ground truth ($p\rightarrow g$) and from ground truth to prediction ($g\rightarrow p$), and their mean (mMAE), as shown in Table~\ref{tab:easycom_asl}. We report the baseline numbers directly from the cited papers. In addition to full model fine-tuning, we also evaluated other fine-tuning approaches: (i) \textit{frozen}, where the M-BEST-RQ encoder is kept frozen and only train the last linear layers are fine-tuned, and (ii) ``weighted comb.'', which additionally uses a weighted combination of all conformer layer outputs with trainable weights. Despite having a much smaller number of trainable parameters, these models outperformed the AV models which use DOA+image and AV+raw-audio as inputs. Nevertheless, the MAE$_{p\rightarrow g}$ was found to be high, indicating that these models are more likely to hallucinate extra speech sources. Full model fine-tuning was able to solve this problem, with the fine-tuned system (D) outperforming all baselines trained on audio-visual (AV) inputs with a state-of-the-art mMAE of 5.6 degrees.

\begin{table}[tb]
    \centering
    \caption{S-ASL results on the EasyCom dataset. We report MAE$_{p\rightarrow g}$ (false positives), MAE$_{g\rightarrow p}$ (missing targets), and their mean mMAE, all in degrees. ``AV'' indicates whether audio-visual input is used in the model. M-BEST-RQ models (C) and (D) are as defined in Table~\ref{tab:mmcsg_wer}.}
    \label{tab:easycom_asl}
    \centering
    \setlength{\tabcolsep}{3pt}
    \adjustbox{max width=\linewidth}{%
    \begin{tabular}[width=\linewidth]{@{} l c r r r r @{}}
        \toprule
        \textbf{Model ($\leftarrow$ Input)} & \textbf{AV} & \textbf{Size~(M)} & MAE$_{p\rightarrow g}$ & MAE$_{g\rightarrow p}$ & mMAE\\
        \midrule
        \cite{jiang2022egocentric} $\leftarrow$ DOA & \xmark & 15.8 & 129.8 & 46.5 & 88.1 \\
        \cite{jiang2022egocentric} $\leftarrow$ AV + cor & \cmark & 28.4 & 16.8  & 6.6 & 11.7 \\
        \cite{jiang2022egocentric} $\leftarrow$ AV + spec & \cmark & 28.4 & 8.8  & 6.2 & 7.5 \\
        \cite{jiang2022egocentric} $\leftarrow$ DOA + image & \cmark & 28.4 & 66.8 & 36.5 & 51.7 \\
        \cite{jiang2022egocentric} $\leftarrow$ AV + raw-audio & \cmark & 28.4 & 40.1  & 140.8 & 90.5 \\
        \cite{zhao2023audio} AVSL (scratch) & \cmark & 10.7 & 9.3 & 4.7 & 7.0 \\
        \cite{zhao2023audio} AVSL (pre-trained) & \cmark & 10.7 & \textbf{8.0}  & \textbf{4.5} & \textbf{6.3} \\
        \midrule
        (C) + frozen & \xmark & 4.2 & 25.9 & 6.4 & 16.2 \\
        \quad ~ + weighted comb. & \xmark & 4.2 & 24.0 & 4.8 & 14.4 \\
        \quad ~ + full fine-tune & \xmark & 99.7 & 4.9 & 7.0 & 6.0 \\
        (D) + frozen & \xmark & 4.2 & 26.7 & 6.3 & 16.5 \\
        \quad ~ + weighted comb. & \xmark & 4.2 & 22.0 & \textbf{4.6} & 13.3 \\
        \quad ~ + full fine-tune & \xmark & 99.7 & \textbf{4.5} & 6.7 & \textbf{5.6} \\
        \bottomrule
    \end{tabular}
    }%
\end{table}

\subsection{Glasses Wearer Voice Activity Detection}

Table~\ref{tab:easycom_vad} shows the mAP numbers for the W-VAD task, with the baselinee numbers reported directly from the cited papers. All baselines were initialized from the AV models trained on the S-ASL task, whereas our models are fine-tuned only on the W-VAD task. We found that our fine-tuned model (C) achieved over 90\% mAP, which is comparable with the baselines. Nevertheless, the best baseline results are obtained using spectrogram features, suggesting that the use of log-Mel features in M-BEST-RQ may be suboptimal. Furthermore, since EasyCom does not provide official mouth-directed ATFs, we designed the mouth beamformer solely based on the array geometry. This may also explain the relatively weaker performance, as the beamformed signal from the wearer's mouth may be a strong indicator for the W-VAD task.

\begin{table}[tb]
    \centering
    \caption{W-VAD results on the EasyCom dataset, in terms of mAP ($\uparrow$). All baselines contain $<$1M trainable params, similar to the ``frozen'' and ``weighted comb.'' versions of our M-BEST-RQ models. ``AV'' indicates whether audio-visual input is used during pre-training.}
    \label{tab:easycom_vad}
    \centering
    \begin{tabular}[width=\linewidth]{@{} l c r@{}}
        \toprule
        \textbf{Model ($\leftarrow$ Input)} & \textbf{AV Pre-train} & \textbf{mAP} \\
        \midrule
        \cite{jiang2022egocentric} $\leftarrow$ cor  & \cmark & 90.20 \\
        \cite{jiang2022egocentric} $\leftarrow$ energy & \cmark & 88.89 \\
        \cite{jiang2022egocentric} $\leftarrow$ spec & \cmark & 91.69 \\
        \cite{jiang2022egocentric} $\leftarrow$ AV + raw-audio & \cmark & 87.29 \\
        \cite{zhao2023audio} AVSL & \cmark & \textbf{93.70} \\
        \midrule
        (C) + frozen & \xmark & 86.66 \\
        \quad ~ + weighted comb. & \xmark & 87.75 \\
        \quad ~ + full fine-tune & \xmark & \textbf{90.16} \\
        (D) + frozen & \xmark & 86.12 \\
        \quad ~ + weighted comb. & \xmark & 87.72 \\
        \quad ~ + full fine-tune & \xmark & 89.29 \\
        \bottomrule
    \end{tabular}
\end{table}

\section{Conclusion}\label{sec:conclusion}
We introduced M-BEST-RQ, the first foundation model designed specifically for smart glasses, and curated three downstream tasks to evaluate its performance: C-ASR, S-ASL, and W-VAD. Evaluations on the MMCSG and EasyCom datasets demonstrated the utility of M-BEST-RQ for multi-channel speech across devices. On the conversational ASR task, M-BEST-RQ fine-tuned with only 8 hours labeled data outperformed strong ASR baselines trained on 2k+ labeled hours. With EasyCom, we showed the cross-device generalizability of M-BEST-RQ, where it matched or outperformed state-of-the-art results on the source localization and wearer VAD tasks. We believe that advancements in these tasks have significant potential to improve the user experience in wearable devices. With this framework in place, future work can build upon our model by exploring different SSL techniques or input features, and developing streaming and lightweight foundation models to facilitate seamless deployment.

\noindent
\textbf{Acknowledgments.} We thank Kate\v{r}ina \v{Z}mol\'{i}kov\'{a}, Morrie Doulaty, Christi Miller, and Calvin Murdock for their help in preparing the pre-training data and downstream tasks.

\bibliographystyle{IEEEbib}
\bibliography{refs}

\end{document}